\pdfoutput=1

\documentclass[aps,prl,twocolumn,showpacs,floatfix,letterpaper,longbibliography]{revtex4-1}
\usepackage{graphicx}
\usepackage[english]{babel}
\usepackage{bm}
\usepackage{amsmath,amsfonts}
\usepackage{amsbsy}
\usepackage[colorlinks=true,linkcolor=blue,urlcolor=blue,citecolor=blue,breaklinks=true]{hyperref}
\usepackage[utf8]{inputenc}
\DeclareUnicodeCharacter{03BC}{\mbox{$\mu$}}
\DeclareUnicodeCharacter{2009}{ }

\usepackage{graphics}
\usepackage{subcaption}
\usepackage{xcolor}
\usepackage{bbm}
\def\bm#1{\mbox{\boldmath{$#1$}}}

\newcommand{\be}{\begin{equation}}
\newcommand{\ee}{\end{equation}}

\usepackage[normalem]{ulem}
\newcommand{\beq}{\begin{equation}}
\newcommand{\eeq}{\end{equation}}
\newcommand{\beqa}{\begin{eqnarray}}
\newcommand{\eeqa}{\end{eqnarray}}
\newcommand{\bea}{\begin{eqnarray}}
\newcommand{\eea}{\end{eqnarray}}
\usepackage{amssymb}
\usepackage{array}
\usepackage{amsmath}
\usepackage{graphicx}\usepackage{graphics}
\usepackage{dcolumn}
\usepackage{bm}\usepackage{varioref}

\begin{document}

\title{Debye mechanism of giant microwave absorption in superconductors}

\author{M. Smith}

\affiliation{Department of Physics, University of Washington, Seattle, WA 98195 USA}

\author{A. V. Andreev}

\affiliation{Department of Physics, University of Washington, Seattle, WA 98195 USA}

\author{B.Z. Spivak}

\affiliation{Department of Physics, University of Washington, Seattle, WA 98195 USA}

\date{July 20, 2019}

\begin{abstract}
We discuss a mechanism  of microwave absorption in conventional superconductors which is similar to the Debye absorption mechanism in molecular gases.
The contribution of this mechanism to the \emph{ac} conductivity is proportional to the inelastic quasiparticle relaxation time $\tau_\mathrm{\mathrm{in}}$ rather than the elastic one $\tau_{\mathrm{el}}$ and
therefore it can be much larger than the conventional one. The Debye contribution to the linear conductivity arises only in the presence of a \emph{dc} supercurrent in the system and its magnitude depends strongly on the orientation of the microwave field relative to the supercurrent. The Debye contribution to the nonlinear conductivity exists even in the absence of \emph{dc} supercurrent. Since it is proportional to $\tau_{\mathrm{in}}$ the nonlinear threshold is anomalously low.
Microwave absorption measurements  may provide  direct information about   $\tau_\mathrm{in}$ in superconductors.
\end{abstract}

\maketitle

\addtolength{\abovedisplayskip}{-1mm}
In this article we discuss the theory of microwave absorption in superconductors. In linear response to the microwave field $\bm{E}(t)=  \bm{E}_\omega \cos (\omega t)$,  and in the limit of low frequencies $\omega$, the current density in a superconductor may be written as
\begin{equation}
\label{eq:current}
{\bf j}=\frac{e}{m} N_\mathrm{s}\, \bm{p}_\mathrm{s}+\sigma {\bf E}.
\end{equation}
Here $N_\mathrm{s}$ is the superfluid density, $e$ and $m$ are, respectively, the charge and the mass of the electron, and the superfluid momentum is defined by $\bm{p}_{s}= \frac{\hbar}{2} \bm{\nabla} \chi - \frac{e}{c}\bm{A}$, with $\chi$ being the  phase of the order parameter, and $\bm{A}$ the vector potential. The second term in Eq.~\eqref{eq:current}, characterized by the conductivity $\sigma$, represents the dissipative part of the current.

The  microwave absorption coefficient is controlled by the conductivity $\sigma$. The value of $\sigma$ is determined by the quasiparticle scattering processes in the superconductor, which are generally characterized by two relaxation times: elastic, $\tau_{\mathrm{el}}$, and inelastic, $\tau_{\mathrm{in}}$, ones.  In a typical situation, which we assume below, $ \tau_{\mathrm{in}} \gg \tau_{\mathrm{el}}$.    The theory of transport phenomena in conventional superconductors was developed long ago, see for example  ~\cite{schrieffer_theory_1999,larkin_nonlinear_1977,aronov_boltzmann-equation_1981,ovchinnikov_electromagnetic_1978,mattis_theory_1958}.
The conventional result is that the conductivity, and consequently the microwave absorption coefficient, are proportional to the elastic relaxation time $\tau_{\mathrm{el}}$.  For example, at temperatures $T$ near the critical temperature $T_{c}$ the conductivity of a superconductor $\sigma$  is \cite{ aronov_boltzmann-equation_1981, ovchinnikov_electromagnetic_1978}
\begin{equation}\label{eq:conventional}
\sigma = \sigma_{\mathrm{D}}\ln \frac{T_c}{\omega},
\end{equation}
where $\sigma_{\mathrm{D}}$  is Drude conductivity of a normal metal~\footnote{Equation~\eqref{eq:conventional} applies at sufficiently high frequencies}.

In this article we discuss another  contribution to
the conductivity, $\sigma_\mathrm{DB}$, that is proportional to the inelastic relaxation time $\tau_{\mathrm{in}}$.
As a result it may exceed the conventional contribution by orders of magnitude.  This contribution to the linear conductivity exists only in the presence of a \emph{dc}  supercurrent. Furthermore, this contribution  is strongly anisotropic and depends on the relative orientation between  $\bm{E}_\omega$  and the supercurrent. Even in situations where this contribution is small in comparison to the conventional result, it determines  the dependence of the conductivity on both the magnitude and direction of the \emph{dc} supercurrent. This enables determination of $\tau_\mathrm{in}$ from measurements of microwave absorption. A contribution to the nonlinear conductivity which is proportional to $\tau_{in}$  exists even at zero \emph{dc} supercurrent. Consequently,  the nonlinear  threshold for microwave absorption turns out to be anomalously low.

The physical mechanism of this contribution to the conductivity is similar to the Debye mechanism of microwave absorption in gases \cite{debye_polar_1970}, Mandelstam-Leontovich mechanism  of the second viscosity in liquids \cite{landau_fluid_2013}, and Pollak-Geballe mechanism of microwave absorption in the hopping conductivity regime \cite{pollak_low-frequency_1961}. It can be qualitatively understood as follows. Let us separate the superfluid momentum $\bm{ p}_\mathrm{s}(t)=\bar{\bm{p}}_\mathrm{s}+\delta \bm { p}_\mathrm{s}(t)$ into a \emph{dc} part $\bar{\bm{p}}_\mathrm{s}$ and the \emph{ac} part $\delta \bm { p}_\mathrm{s}(t)$ whose time evolution is determined by the microwave field
\begin{equation}
  \label{eq:acceleration}
 \delta \dot{\bm{p}}_{s}(t) =e{\bf E} (t).
\end{equation}
At low frequencies, $\omega \ll \tau_{\mathrm{el}}^{-1}$, the quasiparticle distribution function $n$ depends only on the energy $\epsilon$. Importantly, the  density of states per unit energy, $\nu(\epsilon, p_\mathrm{s})$,  depends on the instantaneous value of the superfluid momentum $p_\mathrm{s}$. In other words, as the value of ${\bf p}_{s}$ changes individual quasiparticle levels move in energy space. At finite temperature the quasiparticles occupying these levels travel in energy space as well. This motion creates a non-equilibrium quasiparticle distribution. The relaxation of the latter due to inelastic scattering causes entropy production and energy dissipation.  The corresponding contribution to the conductivity is proportional to $\tau_{in}$.  The reason why the Debye contribution to the linear conductivity exists only at $\bar{\bm{p}}_\mathrm{s} \neq 0$ is the following. The density of states is invariant under time reversal and thus can depend only on  the magnitude of the condensate momentum $p_\mathrm{s} = |\bm{p}_\mathrm{s}|$.
As a result, in the linear in $\bm{E}$ approximation $\nu(\epsilon)$ changes in time proportionally to $\delta \bm{p}_\mathrm{s}(t) \cdot \bar{\bm{p}}_\mathrm{s}$.

To describe the motion of energy levels we note that the number of levels in the system is conserved.  Therefore the density of states $\nu(\epsilon, p_\mathrm{s}(t))$
is subject to the continuity equation in energy space $\partial_t \nu(\epsilon, p_s) +\partial_\epsilon [v_{\nu} (\epsilon)\nu (\epsilon, p_s)] $=0, where $v_\nu (\epsilon, p_s)$ is the level ``velocity'' in energy space. Using the condensate acceleration equation \eqref{eq:acceleration} we can express the latter in the form
$v_{\nu} (\epsilon, p_s)=  e \bm{E} \cdot \bm{V}(\epsilon,\bm{p}_\mathrm s)$, where
\begin{equation}\label{eq:level_velocity}
   \bm{V}(\epsilon,\bm{p}_{\mathrm s}) = - \frac{1}{\nu (\epsilon,p_{\mathrm s})}  \int_{0}^{\epsilon} d \tilde{\epsilon}  \frac{\partial \nu (\tilde{\epsilon}, p_{\mathrm s} )}{\partial \bm{p}_{\mathrm s}}
\end{equation}
characterizes the sensitivity of the energy levels to changes of  ${\bf p}_{s}$. The quasiparticle distribution function $n(\epsilon,t)$ describes the occupancy of energy levels. In the absence of inelastic scattering its time evolution due to the spectral flow is described by the continuity equation $\partial_t(\nu n)+\partial_\epsilon(v_{\nu} \nu n) =0$. Combining this with the continuity equation for $\nu (\epsilon, p_s)$ and allowing for inelastic collisions
we obtain the kinetic equation
\begin{equation}\label{eq:n_dot}
 \partial_{t} n (\epsilon, t)+ e\bm{E}(t)\cdot  \bm{V} (\epsilon, \bm{p}_{\mathrm s})\,  \partial_\epsilon  n(\epsilon, t) = I\{  n\},
\end{equation}
where $I\{ n\}$ is the collision integral describing inelastic  scattering of quasiparticles.

The power $W$ of microwave radiation absorbed per unit volume of the superconductor may be obtained by evaluating the rate of work performed by the microwave field on the quasiparticles, which is given by
\begin{equation}\label{eq:absorption power}
  W = \int_0^\infty d \epsilon \big\langle \nu (\epsilon, p_s (t))n(\epsilon, t)   e\bm{E}(t)\cdot  \bm{V} (\epsilon, \bm{p}_{\mathrm s}(t))\big\rangle.
\end{equation}
Here $\langle \ldots \rangle$ denotes time averaging.  Below we characterize the absorption power by the dissipative part of the conductivity $\sigma_\mathrm{DB} $ defined by
\begin{equation}\label{eq:sigma_W}
  \frac{\sigma_\mathrm{DB}}{2} \,  E_\omega^2 = W.
\end{equation}

\emph{Linear regime.}---For an equilibrium distribution  the integrand in Eq.~\eqref{eq:absorption power}  is a total derivative and $W=0$. At small microwave fields we can linearize the kinetic equation \eqref{eq:n_dot} in ${\bf E}(t)$ and the deviation of the quasiparticle distribution function from equilibrium,  $\delta n(\epsilon, t)=n(\epsilon, t)-n_{F}(\epsilon) \ll 1$ (here  $n_{F}(\epsilon)=[\exp(\epsilon/T)+1]^{-1}$ is  the  Fermi function).
Below we assume that the temperature is near the critical temperature, $|T-T_c|\ll T_c$. In this case
the density of states is affected by the condensate momentum
in a narrow energy window $|\epsilon -\Delta|\ll T$.  Since the energy transfer in a typical inelastic collision is of order $T$ the inelastic collision integral in Eq.~\eqref{eq:n_dot} may be written in the relaxation time approximation,
\begin{equation}\label{eq:I_relaxation_time}
  I\{ n\}= -\frac{\delta n(\epsilon,t)}{\tau_{\mathrm{in}}},
\end{equation}
where the inelastic relaxation time $\tau_{\mathrm{in}}(T)$  depends only on the temperature $T$.

For an isotropic spectrum, which we assume below for simplicity, the vector $\bm{V} (\epsilon,\bm{p}_{\mathrm s})$ is parallel to $\bm{p}_{\mathrm s}$.
In this case only the longitudinal conductivity,  which corresponds to $\bm{E}_\omega \parallel \bar{\bm{p}}_{\mathrm s}$, is affected by inelastic relaxation.
Solving the linearized kinetic equation  in the relaxation time approximation \eqref{eq:I_relaxation_time}, substituting the result into Eq.~\eqref{eq:absorption power}, and using the relation \eqref{eq:sigma_W}  we obtain the ratio of the longitudinal conductivity $\sigma_{\mathrm{DB}}$  to the Drude conductivity $\sigma_\mathrm{D}$ at $|T-T_c|\ll T_c$
\begin{equation}
\label{eq:sigma_ratio}
\frac{\sigma_{\mathrm{DB}}}{\sigma_\mathrm{D}}=
 \frac{3\tau_{\mathrm{in}}}{4\tau_{\mathrm{el}}} \frac{1}{\left[ 1 + \left( \omega \tau_{\mathrm{in}}\right)^2\right]}    \int_0^\infty   \frac{ d \epsilon}{T} \frac{ \nu (\epsilon,\bar{p}_{\mathrm s})  V^2(\epsilon,\bar{p}_{\mathrm s})}{ \nu_n v_{\mathrm F}^2 }.
\end{equation}
Here  $\nu_{n}$ is the normal state density of states at the Fermi level, $v_{\mathrm F}$ is the Fermi velocity,  and  $\sigma_\mathrm{D}= e^2\nu_n D$,  with $D=v_{F}^{2}\tau_{\mathrm{el}}/3$ being  the diffusion coefficient.
Equation \eqref{eq:sigma_ratio} expresses the Debye contribution to the conductivity in terms of the density of states in a current-carrying superconductor. It applies to superconductors with arbitrary symmetry of the order parameter.

Below we focus  on $s$-wave superconductors with an isotropic spectrum and assume  $v_F \bar{p}_\mathrm{s} \ll \Delta$.
In this case the density of states is most strongly affected by the supercurrent at energies near the  gap $\Delta$. Namely at $\bar{p}_\mathrm{s}\neq 0$ the peak in the BCS density of states, $\nu (\epsilon, 0) \to \nu_n \sqrt{\frac{\Delta}{2(\epsilon - \Delta)}}$  at  $\epsilon \to \Delta$, is broadened. The width and the shape of the broadening depends on the magnitude of the condensate momentum $\bar{p}_s$ and the strength of disorder.

\begin{figure}[!t]
    \centering
    \includegraphics[width=0.9\linewidth]{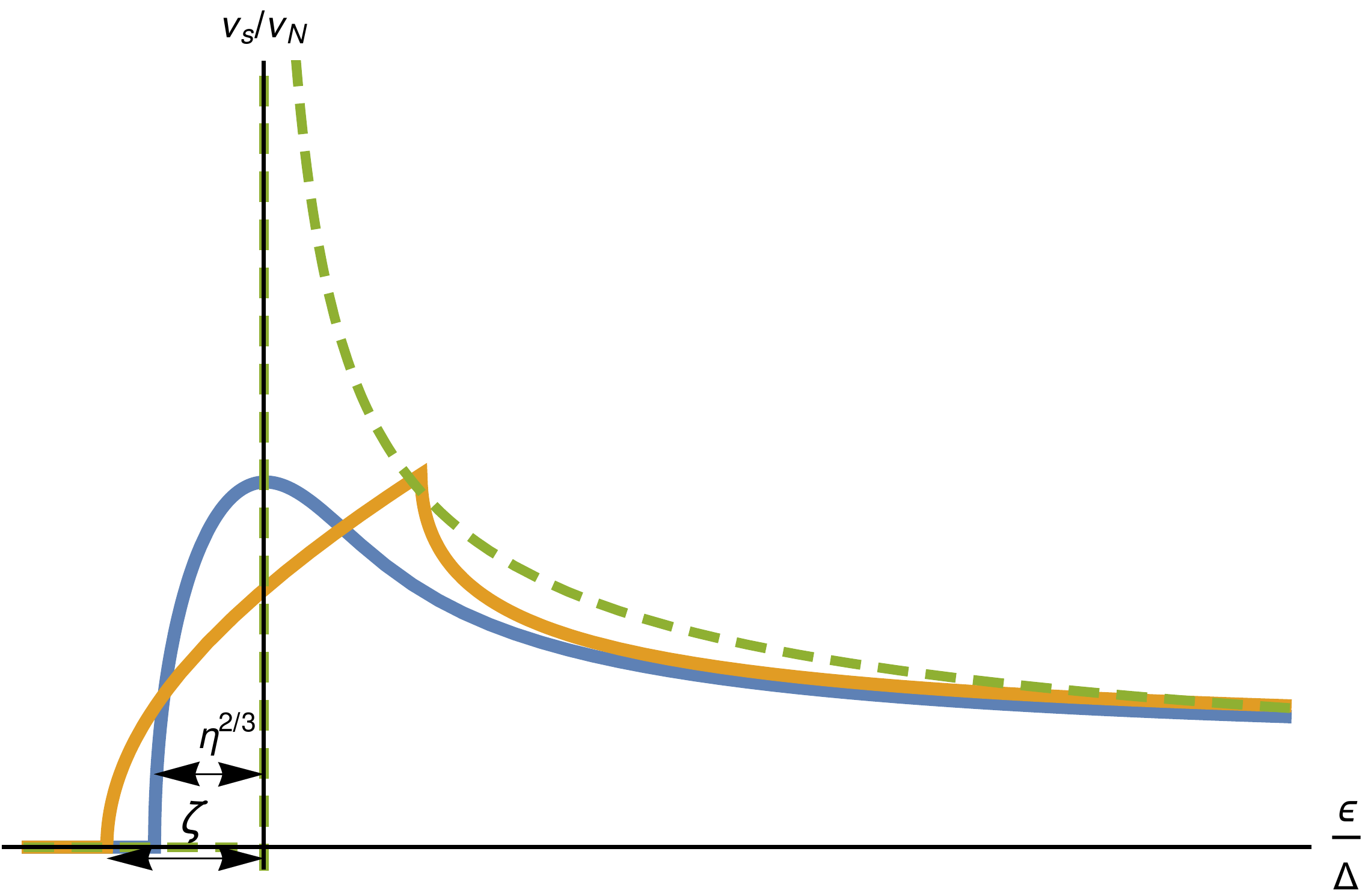}
    \caption{\footnotesize Schematic plots of $\nu (\epsilon, p_\mathrm{s})$ at: $p_\mathrm{s}=0$ - dashed green line, in the diffusive regime $\Delta v_F p_\mathrm{s} \tau_\mathrm{el}^2 \ll 1$ - blue line, and in the ballistic regime $\Delta v_F p_\mathrm{s} \tau_\mathrm{el}^2 \gg 1$ -- orange line.}
    \label{fig:allDOS}
\end{figure}

In the ballistic regime, $v_{\mathrm F} \bar{p}_{\mathrm s}\tau_{\mathrm{el}}^2\Delta \gg 1 $, (which can be realized only in clean superconductors, $\Delta \tau_\mathrm{el} \gg 1$) the density of states can be found from a simple consideration. In this case one can use the standard expression  for the quasiparticle spectrum~\cite{schrieffer_theory_1999, tinkham_introduction_2004,gennes_superconductivity_2018},
$\epsilon( \bm{k} )=\sqrt{ \xi_{\bm{k}}^{2}  +|\Delta|^{2}}+\bm{v}_{\bm{k}} \cdot \bar{\bm{ p}}_\mathrm{s}$, where $\bm{k}$ is the quasiparticle momentum, $\xi_{\bm{k}}$ is the electron energy measured from the Fermi level, and $\bm{v}_{\bm{k}} = d\xi_{\bm{k}}/d \bm{k}$ is the electron velocity. The density of states at $|\epsilon - \Delta | \ll \Delta$ is given by
\begin{equation}\label{eq:purenu}
  \frac{\nu(\epsilon ,  p_{s})}{\nu_n} = \sqrt{\frac{\Delta}{2v_{\mathrm F} p_{\mathrm s}}}\left[\theta(z+1)\sqrt{z+1}  - \theta(z-1)\sqrt{z-1} \right],
\end{equation}
where $z = (\epsilon - \Delta)/v_F p_\mathrm{s} $, and  $\theta (z)$ is the Heavyside step-function.  The width of the broadening of the BCS peak is $\delta \epsilon \sim v_F \bar{p}_\mathrm{s}$. The shape of the broadening is  illustrated in Fig.~\ref{fig:allDOS}. Using Eq.~\eqref{eq:level_velocity}  and  Eq.~\eqref{eq:sigma_ratio}
we obtain for the Debye contribution to the conductivity in the ballistic regime
 \begin{equation}\label{eq:sigma_ballistic}
  \frac{\sigma_\mathrm{DB} }{\sigma_{\mathrm{D}}} =  I_{\mathrm b} \frac{\tau_{\mathrm{in}}}{\tau_{\mathrm{el}}  \left[ 1 + (\omega \tau_\mathrm{in})^2 \right]} \frac{\Delta}{T}  \sqrt{\frac{v_\mathrm{F}  \bar{p}_\mathrm{s} }{\Delta}}
\end{equation}
where $I_{\mathrm b}=\frac{8}{45}$ is a dimensionless integral defined in equation (S.16) of the Supplementary Material~\footnote{See Supplementary Material at (URL) for derivations of Eqs.~\eqref{eq:sigma_ballistic},~\eqref{eq:sigma_diffusive},~\eqref{eq:sigmaNL_balistic}, and~\eqref{eq:sigmaNL_diffusive}.}.
The powerlaw dependence of $\sigma_\mathrm{DB}$ on the condensate momentum $\bar{p}_\mathrm{s}$ follows from the simple scaling form of the density of states in Eq.~\eqref{eq:purenu}.
The exponent of this power law dependence,  $\sigma_\mathrm{DB} \propto \sqrt{\bar{p}_\mathrm{s}}$,  can be understood from the following consideration. The quasiparticle states whose energies are affected by the supercurrent lie in a narrow energy window of width $\delta \epsilon \sim v_\mathrm{F} \bar{p}_\mathrm{s}$ near $\epsilon = \Delta$. The number of such states per unit volume may be estimated as $\nu_n \sqrt{\Delta   v_\mathrm{F} p_\mathrm{s}}$. Since the  characteristic level displacement in the microwave field is given by $   v_\mathrm{F} \delta p_\mathrm{s} \sim v_\mathrm{F} e E_0/\omega$ one obtains an estimate for the absorption power consistent with Eq.~\eqref{eq:sigma_ballistic}.

The above consideration of the density of states is valid  as long the condition of ballistic motion $ v_\mathrm{F} \bar{p}_\mathrm{s} \tau_\mathrm{el} (\epsilon)\gg 1$  is satisfied for
 most of the quasiparticles in the relevant energy interval $|\epsilon - \Delta | \lesssim v_\mathrm{F} \bar{p}_\mathrm{s} $.  Here $\tau_\mathrm{el} (\epsilon)$ is the energy-dependent quasiparticle  mean free time, which for $|\epsilon  -\Delta| \ll \Delta$ is given by the standard expression
 $\tau^{-1}_\mathrm{el} (\epsilon) \approx \tau^{-1}_\mathrm{el}\sqrt{\frac{2(\epsilon - \Delta)}{\Delta}}$  (see for example Ref.~\onlinecite{mineev_introduction_1999}). Therefore the regime of ballistic motion of quasiparticles participating in the Debye mechanism of microwave absorption is realized at relatively  large supercurrent densities, where
 $ v_\mathrm{F} \bar{p}_\mathrm{s}\tau_{\mathrm{el}}^2\Delta \gg 1 $.

To study the Debye contribution to the conductivity outside the ballistic regime we express the density of states in terms of the disorder-averaged Green's functions. This enables us to utilize the standard theoretical methods developed in the theory of disordered superconductors~\cite{larkin_nonlinear_1977, parks_superconductivity:_1969}. We show in the Supplementary Material that the density of states can be expressed as
\begin{equation}
\label{eq:IntermediateDOS}
    \frac{\nu(\epsilon,p_{\mathrm s})}{\nu_n} = \frac{1}{\sqrt{2}}\Im x^{-1},
\end{equation}
where the variable $x$ satisfies the quintic equation (S.9a). In the ballistic regime, $v_\mathrm{F} p_\mathrm{s}\tau_{\mathrm{el}}^2\Delta \gg 1$, the latter reduces to the biquadratic equation (S.11) whose solution, when substituted into Eq.~\eqref{eq:IntermediateDOS}, reproduces Eq.~\eqref{eq:purenu}. In the opposite regime
$v_\mathrm{F} p_\mathrm{s}\tau_{\mathrm{el}}^2\Delta \ll 1$, which corresponds to diffusive motion of quasiparticles in the relevant energy interval, the quintic equation (S.9a) simplifies to
\begin{equation}
\label{eq:Cubic Eqn}
    x\left(x^2+w\right) +\frac{\sqrt{2}\zeta^2}{3 \gamma} = 0.
\end{equation}
Here $\zeta = v_{\mathrm{F}} p_{\mathrm{s}}/\Delta$, $\gamma = (\tau_{\mathrm{el}} \Delta)^{-1}$, and $w=(\epsilon-\Delta)/\Delta$. The solutions of this equation can be written in the scaling form $x= \frac{\zeta^{2/3}}{\gamma^{1/3}} \, \tilde{x} \left( \frac{w \gamma^{2/3}}{\zeta^{4/3}} \right) $, where the explicit form of  $\tilde{x} \left( \frac{w \gamma^{2/3}}{\zeta^{4/3}} \right)$ is given by the Cardano formula, Eq. (S.17).  Substituting this form
into Eq.~\eqref{eq:IntermediateDOS} (the corresponding $\nu(\epsilon, p_\mathrm{s})$ is plotted in Fig.~\ref{fig:allDOS}), and using Eqs.~\eqref{eq:level_velocity} and \eqref{eq:sigma_ratio}, we obtain
\begin{equation}\label{eq:sigma_diffusive}
  \frac{\sigma_\mathrm{DB} }{\sigma_{\mathrm{D}}} = I_{\mathrm d} \frac{\tau_{\mathrm{in}}}{\tau_\mathrm{el}} \frac{\Delta}{T} \frac{\tau_\mathrm{el}\left(\Delta D^2 \bar{p}_{\mathrm s}^4 \right)^{1/3}}{1 + \left( \omega \tau_{\mathrm{in}}\right)^2},
\end{equation}
where $I_{\mathrm d} \approx  0.0549$ is defined in Eq. (S.22) in the Supplementary Material. This expression is consistent with the result obtained in Ref.~\cite{ovchinnikov_electromagnetic_1978} by a different method.

The exponent of the powerlaw dependence $\sigma_{\mathrm DB} \propto \bar{p}_\mathrm{s}^{4/3}$ in Eq.~\eqref{eq:sigma_diffusive} and its order of magnitude  can be obtained by noting that the width of the broadening of the BCS singularity in the diffusive regime is $\delta \epsilon \sim (\Delta D^2 \bar{p}_\mathrm{s}^4)^{1/3}$ and the number per unit volume of levels that participate in microwave absorption is $\sim \nu_n \left( \Delta^2 D \bar{p}_\mathrm{s}^2\right)^{1/3} $.

It is worth noting that the diffusive regime can be realized in both clean, $\Delta \tau_\mathrm{el} \ll 1$, and dirty $\Delta \tau_\mathrm{el} \gg 1$, superconductors. Accordingly Eqs.~\eqref{eq:IntermediateDOS} and \eqref{eq:Cubic Eqn} for the density of states and the resulting conductivity \eqref{eq:sigma_diffusive} can be obtained using either the Gorkov equations or the Usadel equation, see Supplementary Material.

\emph{Nonlinear regime.} Let us now consider the situation in which the \emph{dc} supercurrent is absent, $\bar{p}_\mathrm{s}=0$. In the presence of the microwave field the oscillation amplitude of the condensate momentum is given by $\delta p_\mathrm{s}= eE_\omega/\omega$. Since the Debye contribution to the nonlinear conductivity defined by Eq.~\eqref{eq:sigma_W}  is proportional to $\tau_{in}$  the nonlinear threshold for the microwave absorption is anomalously low. To evaluate microwave absorption in the nonlinear regime it is convenient to introduce the integrated over energy density of states
\begin{equation}\label{eq:N}
  N(\epsilon, t) = \int_0^\epsilon d \varepsilon\ \nu (\varepsilon, t)
\end{equation}
and consider the quasiparticle distribution function not as a function of energy $\epsilon$ and time $t$ but rather as a function of $N$ and $t$.  The change of variables $n(\epsilon, t) \to n(N,t)$ is equivalent to the transformation from Eulerian to Lagrangian variables in fluid mechanics~\cite{landau_fluid_2013}. In this representation the kinetic equation \eqref{eq:n_dot} acquires a very simple form,
\begin{equation}
\label{eq:kineq_Lagrange}
\partial_t n(N,t) = I\{n\} = -\frac{n(N,t) - n_F(\epsilon (N,t))}{\tau_\mathrm{in}}.
\end{equation}
Note that the electric field enters this equation only via the time-dependence of the quasiparticle energy level $\epsilon (N,t)$.
In the presence of the microwave field the latter undergoes nonlinear oscillations $\epsilon (N,t) = \epsilon_0 (N) + \delta \epsilon (N,t) $ whose form is determined by Eq.~\eqref{eq:N}. Note that the linearization of the collision integral is justified in the nonlinear regime as long as the amplitude of $\delta \epsilon (N,t)$ is small as compared to $T$. The solution of Eq.~\eqref{eq:kineq_Lagrange} can be written as
$n(N,t) = n_F (\epsilon_0(N)) + \frac{d n_F (\epsilon_0 (N) )}{d\epsilon_0 (N)}\int_{0}^{\infty} \frac{d\tau}{\tau_\mathrm{in}} e^{- \frac{\tau}{\tau_\mathrm{in}}} \delta \epsilon (N, t-\tau)  $.
The absorption  power per unit volume in this representation is given by
$    W= \int_{0}^{\infty} d N \left \langle  n(N,t) \partial_t \epsilon (N,t) \right\rangle$. Substituting the solution for $n(N,t)$ into this expression we get
\begin{equation}\label{eq:W_N}
  W= \int_{0}^{\infty} d N \int_0^\infty  \frac{d \tau}{4T} e^{- \frac{\tau}{\tau_\mathrm{in}}} \langle \partial_t \epsilon (N, t) \partial_t \epsilon (N, t -\tau) \rangle.
\end{equation}
Here the level velocity is given by  $\partial_t \epsilon (N,t) = v_\nu (\epsilon,t) = e \bm{E}\cdot \bm{V} (\epsilon, \bm{p}_\mathrm{s})$, with $\bm{V} (\epsilon, \bm{p}_\mathrm{s})$  defined in Eq.~\eqref{eq:level_velocity}.  This can be shown by taking the time derivative of \eqref{eq:N}.

Equation \eqref{eq:W_N} expresses the power of nonlinear microwave absorption in terms of the correlation function of level velocities $\partial_t \epsilon (N, t)$, which are defined by Eq.~\eqref{eq:N}. Similarly to the linear regime, the results depend of the degree of disorder, the amplitude of the microwave field $E_\omega$ and the frequency of radiation $\omega$. They simplify in the ballistic regime $e E_\omega v_\mathrm{F} \Delta \tau_\mathrm{el}^2 \gg \omega$ and in the diffusive regime $e E_\omega v_\mathrm{F} \Delta \tau_\mathrm{el}^2 \ll \omega$ where the nonlinear conductivity has a simple powerlaw dependence on the amplitude of the microwave field $E_\omega$. In the ballistic regime we obtain
\begin{equation}\label{eq:sigmaNL_balistic}
\frac{\sigma^\mathrm{nl}_\mathrm{DB}}{\sigma_\mathrm{D}}  =
\frac{\tau_\mathrm{in}}{\tau_\mathrm{el}}\frac{\Delta}{T} \sqrt{\frac{v_\mathrm{F} eE_\omega}{  \omega\Delta} }   \,  F_\mathrm{b}(\omega \tau_\mathrm{in}),
\end{equation}
while in the diffusive regime we find
\begin{equation}\label{eq:sigmaNL_diffusive}
\frac{\sigma^\mathrm{nl}_\mathrm{DB}}{\sigma_\mathrm{D}}  =
\frac{\tau_\mathrm{in}}{\tau_\mathrm{el}}\frac{\Delta}{T}
\frac{\Delta^{1/3} D^{5/3} |e E_\omega|^{4/3}}{v_F^2 \omega^{4/3}} \,  F_\mathrm{d}(\omega \tau_\mathrm{in}).
\end{equation}
The functions $F_\mathrm{b}(\omega \tau_\mathrm{in})$ and  $F_\mathrm{d}(\omega \tau_\mathrm{in})$ that describe  the frequency dependence of nonlinear microwave conductivity are given Eqs. (S.35) and (S.41) in the Supplementary Material.
Although, in  contrast to Eqs.~\eqref{eq:sigma_ballistic} and \eqref{eq:sigma_diffusive}, they do not have a simple Lorentzian form, their high- and low-frequency asymptotic behavior is similar;
at low frequency $F_\mathrm{b}(0)  \approx 0.10848$ and $F_\mathrm{d}(0)  \approx 0.10909 $ while at high frequencies they behave as $1/(\omega \tau_\mathrm{in})^2$.

The nonanalytic dependences of $\sigma_\mathrm{DB}$ on $\bar{p}_\mathrm{s}$ in Eqs.~\eqref{eq:sigma_ballistic} and
\eqref{eq:sigma_diffusive}, and of $\sigma_\mathrm{DB}^{\mathrm{nl}}$ on $E_\omega$  and $\omega$  in Eqs.~\eqref{eq:sigmaNL_balistic} and \eqref{eq:sigmaNL_diffusive}
are related to the divergence of the BCS density of states at $\epsilon = \Delta$.
In real superconductors this divergence is smeared by the anisotropy of the order parameter in ${\bf k}$ space and by  pairbreaking processes, which are characterized by a broadening parameter $\Gamma \ll | \Delta |$. Consequently,  at $\bar{p}_\mathrm{s} \to 0$ the $\bar{p}_\mathrm{s}$-dependence of the conductivity should become analytic, $\sigma_{DB} = c \,  \bar{p}_\mathrm{s}^2$.
The manitude of the coefficient $c$ can be estimated by matching this expression to Eqs.~\eqref{eq:sigma_ballistic} and \eqref{eq:sigma_diffusive} at the values of $\bar{p}_\mathrm{s}$  determined by the condition that  the  energy broadening of the BCS singularity, $\delta \epsilon (\bar{p}_\mathrm{s})$ be of order   $\Gamma$.

So far we focused our consideration on  the temperature interval near $T_{c}$.
At low temperatures, $T\ll \Delta$, the dimensionless quasiparticle concentration  $x=(\nu_n \Delta)^{-1}\int_{0}^{\infty} d \epsilon \nu (\epsilon ) n_{F}(\epsilon)  $  in \emph{s}-wave superconductors is exponentially small;  $x\sim \sqrt{ \frac{T}{\Delta}} \exp(-\Delta/T)$ . Consequently the conventional contribution to the microwave absorption coefficient, which is  proportional to $\tau_\mathrm{el}$,  is exponentially small as well. It is interesting that the Debye contribution to the microwave absorption coefficient in this regime does not contain the exponentially small factor and is, roughly speaking, comparable to that near $T_{c}$.
Indeed, at $T\ll \Delta$ there are two  quasiparticle inelastic relaxation rates in superconductors. The quasiparticle-phonon relaxation processes that conserve the number of quasiparticles are characterized by the rate $1/\tau_{\mathrm{in}}^{(st)}(T)$ (which is of the same order as a electron-phonon relaxation rate in normal metals, see for example \cite{aronov_boltzmann-equation_1981}). The second type of inelastic relaxation processes correspond to  recombination, which changes the total number of quasiparticles. The recombination rate is proportional to the quasiparticle concentration $1/\tau_{r}\sim x/ \tau_{r}^{(0)}\ll \tau_{\mathrm{in}}^{(st)}$, where $\tau_{r}^{(0)}\sim \tau_{\mathrm{in}}^{(st)}(\Delta)$.  The Debye contribution to the dissipative kinetic coefficients  is proportional to the longest relaxation time in a system (see for example \cite{landau_fluid_2013}. In our case  it is $\tau_{r}$. Therefore, in the low frequency limit, $\omega\tau_{r}\ll 1$, the exponentially small factor $\exp(-\Delta/T)$ is canceled from the expression for the conductivity. Below we illustrate this fact in the diffusive regime, and  at $\Delta\gg \delta \epsilon(p_{s})\gg T$.  In this case the magnitude of  the level velocity in the interval of energies of order  $T$ is $ V \sim \frac{1}{\bar{p}_\mathrm{s}}  \delta \epsilon \sim   \left( \Delta D^2 \bar{p}_\mathrm{s} \right)^{1/3} $ and we get from Eq.~\eqref{eq:sigma_ratio}
\begin{equation}
\frac{\sigma_\mathrm{DB}}{\sigma_\mathrm{D}}\sim \frac{\tau_{r}^{(0)}}{\tau_{\mathrm{el}}}  \, 
\left[ \left(\Delta \tau_{el}\right) \, \left(v_{F}p_{s}\tau_{\mathrm{el}} \right) \right]^{2/3}.
\end{equation}
We would like to note that at very low temperatures the microwave absorption in gapped superconductors  is controlled by two level systems (see for example Ref.~\onlinecite{zhang_microresonators_2019} and references therein).

The situation with a spatially uniform supercurrent density that was considered above 
 can only be realized  in thin superconducting films.
In bulk superconductors in the presence of a magnetic field  $ H < H_{c1}$ that is parallel to the surface $\bar{p}_\mathrm{s}$ is nonzero only within the London  penetration depth  $\lambda_{L}$ near the surface.  The mechanism of microwave absorption discussed above will still apply in this geometry and the presented above results still hold up to a numerical factor of order unity.  The reason for this is that the quasiparticles that give the main contribution to microwave absorption have energies that lie in a narrow interval near the gap, $|\epsilon - \Delta| \lesssim \delta \epsilon$, where $\delta \epsilon = v_\mathrm{F} \bar{p}_\mathrm{s}$  in the ballistic regime and $\delta \epsilon = \left( \Delta D^2  \bar{p}_\mathrm{s}^4\right)^{1/3}$ in the diffusive regime.
Roughly half of these quasiparticles have energies below $\Delta$ and therefore they are trapped near the surface within a distance of order $\lambda_{L}$.

The  microwave absorption coefficient in thin films of \emph{d}-wave superconductors should  be also proportional to $\tau_{\mathrm{in}}$.  However, its  dependence on $\bar{p}_\mathrm{s}$  is expected to be different from those in Eqs.~\eqref{eq:sigma_ballistic}, \eqref{eq:sigma_diffusive}, \eqref{eq:sigmaNL_balistic}, and \eqref{eq:sigmaNL_diffusive}.  Moreover,  in bulk samples of $d$-wave superconductors in the presence of a magnetic field parallel to the surface the situation is substantially different. In a gapless superconductor the quasiparticles in the relevant energy interval can diffuse into the bulk. Therefore in this case the characteristic relaxation time is given by the minimum between the inelastic relaxation time and the time of diffusion from the surface layer of thickness $\lambda_{L}$.

Finally we would like to note that the considered above mechanism of the microwave absorption is closely related to the mechanism of \emph{ac} conductivity of SNS junctions discussed in Ref.~\onlinecite{zhou_resistance_1997}.

\acknowledgments

\begin{acknowledgments}

This work was supported by the U.S.  Department of Energy Office
of Science, Basic Energy Sciences under Award No. DE-FG02-07ER46452 and by the NSF grant  MRSEC  DMR-1719797.

\end{acknowledgments}

\end{document}


\title{Supplementary Material to Debye mechanism of giant microwave absorption in superconductors}

\author{M. Smith}

\affiliation{Department of Physics, University of Washington, Seattle, WA 98195 USA}

\author{A. V. Andreev}

\affiliation{Department of Physics, University of Washington, Seattle, WA 98195 USA}

\author{B.Z. Spivak}

\affiliation{Department of Physics, University of Washington, Seattle, WA 98195 USA}

\date{July 20, 2019}

\maketitle


\section{Derivation of the Debye contribution to the linear conductivity}

In order to evaluate the linear conductivity in the presence of supercurrent we express the  density of states in a superconductor in terms of the dimensionless disorder-averaged retarded Green's function  $g(\epsilon)$  at coinciding points
\begin{equation}
  \label{eq:DOSGfrel}
\frac{\nu(\epsilon,p_\mathrm{s})}{\nu_n} = -\frac{2}{\pi} \mathrm{Im} \,  g (\epsilon ).
\end{equation}
The latter can be expressed as~\cite{abrikosov_methods_1975, parks_superconductivity:_1969}
\begin{eqnarray}\label{eq:g_def}
  g (\epsilon) &=& \left\langle \frac{\tilde{\epsilon} - \bm{v}\cdot \bm{p}_{\mathrm s}}{\sqrt{( \tilde{\epsilon}- \bm{v}\cdot \bm{p}_{\mathrm s})^2 - |\tilde{\Delta} |^2 }} \right\rangle,
\end{eqnarray}
where $\langle \ldots \rangle$ denotes averaging over the Fermi surface and the disorder-renormalized energy $\tilde{\epsilon}$ and order parameter $\tilde{\Delta}$ are given by
\begin{subequations}\label{eq:epsilon_Delta_tilde}
  \begin{eqnarray}\label{eq:epsilon_tilde}
  \tilde{\epsilon} &=&  \epsilon + \frac{i}{2\tau_{\mathrm{el}} }
  \left\langle \frac{\tilde{\epsilon} - \bm{v}\cdot \bm{p}_{\mathrm s}}{\sqrt{( \tilde{\epsilon}- \bm{v}\cdot \bm{p}_{\mathrm s})^2 - |\tilde{\Delta} |^2 }} \right\rangle
   ,   \\
  \label{eq:Delta_tilde}
 \tilde{\Delta} & = &  \Delta + \frac{i}{2\tau_{\mathrm{el}}}
  \left\langle \frac{ \tilde{\Delta}}{\sqrt{( \tilde{\epsilon}- \bm{v}\cdot \bm{p}_{\mathrm s})^2 - |\tilde{\Delta} |^2 }}\right\rangle,
\end{eqnarray}
\end{subequations}
where we set $\hbar = 1$. For simplicity we assume the Fermi surface to be spherical.  Performing
the angular averaging and introducing the dimensionless variables $\delta = \tilde{\Delta}/\Delta$, $u=\tilde{\epsilon}/\tilde{\Delta}$ we can write the disorder-renormalization equations   for $\tilde{\epsilon}$ and $\tilde{\Delta}$  in the following form
\begin{subequations}\label{eq:renormalization_system}
  \begin{eqnarray}\label{eq:renormalization_epsilon}
         u \delta  &=& \frac{ \epsilon}{\Delta} + \frac{i \delta \gamma}{4\zeta} \left(\sqrt{ u_+^2 - 1} - \sqrt{u_-^2 - 1} \right),  \\
         \label{eq:renormalization_delta}
       \delta &=& 1 + \frac{i\delta \gamma}{4 \zeta}\ln\left(\frac{ u_- - \sqrt{ u_-^2 - 1}}{u_+ - \sqrt{u_+^2 - 1}} \right).
\end{eqnarray}
\end{subequations}
Here  $\gamma = 1/(\tau_{\mathrm{el}}\Delta)$ characterizes the disorder strength,   $\zeta = v_{\mathrm F} p_{\mathrm s}/\Delta$ is a dimensionless measure of the condensate momentum, and we introduced  $u_\pm = u \pm \zeta/\delta$. The Green's function $g(\epsilon)$ in Eqs.~\eqref{eq:DOSGfrel} and \eqref{eq:g_def} is expressed in terms of these variables as
\begin{equation}\label{eq:g_u_delta}
  g = \frac{2i}{\gamma}\left(\frac{\epsilon}{\Delta}-u\delta \right).
\end{equation}
The system of Eqs.~\eqref{eq:renormalization_system}, \eqref{eq:g_u_delta} and \eqref{eq:DOSGfrel} describes the energy dependence of the density of states in the presence of supercurrent.

In the limit $\zeta \to 0$ the solution of Eqs.~\eqref{eq:renormalization_system} is given by  $u_0=\epsilon/\Delta$, and  $\delta_0= 1 + \frac{i\gamma}{2\sqrt{u^2 -1}}$. When substituted into Eqs.   \eqref{eq:DOSGfrel} and  \eqref{eq:g_u_delta} this yields the conventional result for the density of states, $\nu (\epsilon)/ \nu_n =  \epsilon /\sqrt{\epsilon^2 - \Delta^2}$.

At  $\zeta =v_{\mathrm F} p_{\mathrm s}/\Delta \ll 1$ the density of sates  $\nu(\epsilon)$
is significantly affected by the condensate momentum $p_{\mathrm s} \neq 0$ only for energies near the gap, $| \epsilon - \Delta | \ll |\Delta |$, see Eq.~(10). This interval of energies corresponds to $|u_
\pm -1| \ll 1$. The  $p_{\mathrm s}$  dependence of $\nu(\epsilon,p_{\mathrm s})$  in this interval is nonanalytic, c.f. Eq. (10),  and
the  expansion of Eqs.~\eqref{eq:renormalization_system} in powers of $u_+-u_- =2\zeta/\delta$  fails.
Formally, the expansion fails  because at $\zeta=0$
the  solution of Eq.~\eqref{eq:renormalization_system}   approaches the branching point, $u=1$, of the radicals in that equation.
To circumvent this difficulty we introduce new variables $x$ and $y$ via
\begin{equation}
\label{eq:rational}
u=1-x^2 -y^2, \quad \delta = -\frac{ \zeta }{2 xy}
\end{equation}
and rewrite Eqs.~\eqref{eq:renormalization_system} in the form
\begin{widetext}
\begin{subequations}
\label{eq:renormalization_xy}
\begin{eqnarray}
     -\frac{\zeta(1-x^2-y^2)}{2xy} &=& 1+w - \frac{\gamma}{4\zeta d}\left[(x + y)\sqrt{2 - (x + y)^2} - (x - y) \sqrt{2 - (x - y)^2} \right],\\
    -\frac{\zeta}{2xy} &=& 1+\frac{i\gamma}{4\zeta d}\ln\left[\frac{1 - (x - y)^2 -i (x - y)\sqrt{2- (x - y)^2}}{1-(x + y)^2 -i (x + y) \sqrt{2- (x + y)^2}} \right].
\end{eqnarray}
\end{subequations}
\end{widetext}
Here we have introduced the notation $w=(\epsilon -\Delta)/\Delta$.

The advantage of writing the disorder renormalization equations in terms of the variables $x$ and $y$ is that the branching points of the radicals in Eqs.~\eqref{eq:renormalization_system}, which are located at $u_\pm =1-(x\pm y)^2=1$, are resolved in terms of rational functions of $x$ and $y$. As a result the expressions in the right hand side in Eqs.~\eqref{eq:renormalization_xy} become amenable to a series expansion in $x$ and $y$.

Although  $x$ and $y$ enter Eqs.~\eqref{eq:rational} and \eqref{eq:renormalization_xy}  on equal footing we choose $y$ to be odd in $\zeta$ so that in the absence of supercurrent $y=0$.
At small supercurrent, $\zeta \ll 1$, the variable $y$ is also small. In this regime,  the density of states in the relevant energy interval near the spectrum edge may be determined by expanding the expressions in the right hand side of Eqs.~\eqref{eq:renormalization_xy}  to third order in $y$:
\begin{widetext}
\begin{subequations}
\label{eq:YExpandEqns}
\begin{eqnarray}
    \frac{\zeta}{\gamma}\left(1-x^2-y^2+ \frac{2xy(1+w)}{\zeta} \right) &=& \frac{y}{\sqrt{2-x^2}}\left[x^2 -1 + \frac{y^2}{(2-x^2)^2} \right],\\
    -\frac{\zeta}{\gamma}\left(\frac{2xy}{\zeta}+1 \right) &=& \frac{y}{\sqrt{2-x^2}}\left[1+\frac{1}{3}\frac{(1+x^2)y^2}{(2-x^2)^2} \right].
\end{eqnarray}
\end{subequations}
\end{widetext}

The relevant energy interval,  $|w| \ll 1$, corresponds to $|x|\ll 1$. In this case Eqs.~\eqref{eq:YExpandEqns} can be simplified to
\begin{subequations}
\label{eq:IntermediateEqns}
\begin{eqnarray}\label{eq:IntermediateX}
    x \! \left(x^2+w \right) \! \!\left( 2  x+\frac{\gamma}{\sqrt{2}} \right)^2  \!&=& -\zeta^2\! \left(x +\frac{\gamma}{3 \sqrt{2}} \right), \\ \label{eq:IntermediateY}
    y  &=& - \frac{ \sqrt{2} \zeta}{2\sqrt{2} x + \gamma }.
\end{eqnarray}
\end{subequations}
The density of states is obtained by substituting Eqs.~\eqref{eq:rational} into \eqref{eq:g_u_delta} and expanding in small $y$ then $x$.
Within the accuracy of our approximation $\nu(\epsilon, p_{\mathrm s})$ in the relevant energy interval, $|w | \ll 1$,  is given by
\begin{equation}
\label{eq:IntermediateDOSA}
    \nu(\epsilon,p_{\mathrm s}) = \frac{\nu_n}{\sqrt{2}}\Im x^{-1},
\end{equation}
and may be determined by solving Eq.~\eqref{eq:IntermediateX}.

The quintic equation \eqref{eq:IntermediateX} for the variable $x$ has five roots. The complex solutions come in pairs of complex conjugate numbers corresponding to retarded and advanced Green's functions. The retarded solutions correspond to $\Im x^{-1} \geq 0$, c.f. Eq.~\eqref{eq:IntermediateDOSA}. The roots of a general quintic equation can be expressed via the Jacobi theta functions, however Eq.~\eqref{eq:IntermediateX} simplifies significantly in the limiting regimes $ \zeta /\gamma^2 =  v_\mathrm{F} \bar{p}_\mathrm{s}\tau_{\mathrm{el}}^2\Delta  \gg 1 $, and
$  \zeta /\gamma^2 = v_\mathrm{F} \bar{p}_\mathrm{s}\tau_{\mathrm{el}}^2\Delta \ll 1 $, which correspond to, respectively, ballistic and diffusive motion of quasiparticles that participate in microwave absorption.

\subsection{Ballistic Regime}

At relatively large supercurrent densities,  $\zeta/\gamma^2 \gg 1$,  the relevant root of Eq.~\eqref{eq:IntermediateX}  satisfies the condition $|x|\gg \gamma$. In this case we may neglect $\gamma$ in Eq.~\eqref{eq:IntermediateX} to get a quadratic equation for $x^2$:
\begin{eqnarray}
   x^4 + w x^2 + \frac{\zeta^2}{4} = 0,
\end{eqnarray}
which yields the solution
\begin{eqnarray}
    x^2 = \frac{\zeta^2}{\left(\sqrt{-w-\zeta}-\sqrt{-w+\zeta} \right)^2}.
\end{eqnarray}
Substituting this into Eq.~\eqref{eq:IntermediateDOSA}
we obtain Eq.~(10) in the main text;
\begin{equation}
\label{eq:BallisticDOS}
    \frac{\nu(\epsilon,p_\mathrm{s})}{\nu_n} = \frac{1}{\sqrt{2 \zeta}}
    \left[
\eta_b\left(z+1\right) - \eta_b\left(z-1\right)
    \right].
\end{equation}
Here the dimensionless energy variable $z$ and the function $\eta_b(x)$ are defined by
\begin{equation}
\label{eq:z_eta_b_def}
  z= \frac{w}{\zeta}, \quad  \eta_b(x) = \theta(x) \sqrt{x}.
\end{equation}
Substituting Eq.~\eqref{eq:BallisticDOS} into  Eq.~(4)  we obtain,
\begin{eqnarray}
\label{eq:CleanLevelVelocity}
    \frac{V(\epsilon,p_s)}{v_F}&=& \frac{\eta_b(z+1) + \eta_b(z-1) }{\eta_b (z+1) - \eta_b(z-1) }
     \nonumber \\
    &-& \frac{2}{3}\frac{\eta^3_b\left(z+1\right)  - \eta^3_b\left(z-1\right) }{\eta_b\left(z+1\right)  - \eta_b\left(z-1\right) }.
\end{eqnarray}
Substituting Eq.~\eqref{eq:BallisticDOS} into Eq.~(9) and using the dimensionless variable $z$ we obtain Eq.~(11) for the longitudinal conductivity with $I_b$ given by
\begin{equation}
\label{eq:Ib}
    I_b = \frac{3}{4} \int_{-1}^\infty dz  \frac{ \left[\eta_b\left(z+1\right) - \eta_b\left(z-1\right) \right]}{\sqrt{2}} \frac{V^2(\epsilon, p_s)}{v_F^2}.
\end{equation}
Substituting Eq.~\eqref{eq:CleanLevelVelocity} into Eq.~\eqref{eq:Ib} we obtain
\begin{equation}
	I_b = \frac{8}{45}.
\end{equation}

\subsection{Diffusive Regime}

At small supercurrent densities,  $\zeta/\gamma^2 \ll 1$, the relevant root of Eq.~\eqref{eq:IntermediateX} satisfies the condition $|x|\ll \gamma$. In this regime,
 which corresponds to diffusive motion of quasiparticles participating in microwave absorption, Eq.~\eqref{eq:IntermediateX} simplifies to the cubic equation
\begin{equation}
\label{eq:CubicEqnA}
    x\left(x^2+w\right) +\frac{\sqrt{2}\zeta^2}{3 \gamma} = 0.
\end{equation}
Using the Cardano formula and substituting the root with  $\Im x^{-1} \geq 0$ into
 Eq.~\eqref{eq:IntermediateDOSA} we
can express the density of states  in terms of rescaled variables $\eta \equiv \frac{2\zeta^2}{3\gamma}$, and  $\tilde w = w \eta^{-2/3}$ in the form
\begin{eqnarray}
\label{eq:UsadelDOS3}
    \nu(\epsilon,p_s)\!&=\!& \nu_n\frac{\tilde \nu_d(\tilde w)}{\eta^{1/3}} ,\\
    \label{eq:nutildedDef}
    \tilde \nu_d(\tilde w)\!& = \!& \frac{1}{2\sqrt{3}}\theta\left(\tilde w + \frac{3}{2}\right) \!\! \left[\frac{\tilde\alpha(\tilde w)}{2^{2/3}}-\frac{2^{2/3}w^2}{\tilde\alpha(\tilde w)} \right]  ,
\end{eqnarray}
where the function $\tilde\alpha(\tilde w) $ is defined by
\begin{equation}
    \tilde\alpha(\tilde w) = \left(4 \tilde{w}^3 + 27  + \sqrt{27} \sqrt{8\tilde{w}^3 + 27} \right)^{1/3}.
\end{equation}
Substituting this form into  Eq.~(4) we obtain
\begin{equation}
\label{eq:UsadelLevelVelocity}
    V(\epsilon,p_s) = \frac{1}{\tilde\nu_d(\tilde{w})}\frac{Dp_s}{\eta^{1/3}}I(\tilde w),
\end{equation}
where $D$ is the diffusion coefficient and $I(\tilde w)$ is given by
\begin{eqnarray}
I(\tilde{w}) &=& \int_{-\frac{3}{2}}^{\tilde{w}} \frac{d\tilde{x}}{2^{5/3}\sqrt{3}}\left[\frac{2^{4/3}\tilde{x}^2}{\tilde{\alpha}(\tilde{x})}-\tilde{\alpha}(\tilde{x})+ \right. \nonumber \\
&&\left.
\left(\frac{1}{  \tilde{\alpha}^2(\tilde{x})}+  \frac{2^{4/3}\tilde{x}^2}{\tilde{\alpha}^4(\tilde{x})} \right)\left(18 + \frac{\sqrt{3}}{2}\left[\frac{16\tilde{x}^3+ 108}{\sqrt{8\tilde{x}^3+27}} \right] \right)   \right]. \nonumber
\end{eqnarray}
Using Eqs.~\eqref{eq:UsadelLevelVelocity} and ~\eqref{eq:UsadelDOS3} in Eq.~(9) we get Eq.~(14)  for the longitudinal Debye conductivity where the definite integral $I_d$ is given by
\begin{eqnarray}
\label{eq:Id}
   I_d &=&  \frac{1}{2^{5/3}}\int_{-3/2}^\infty d\tilde x \frac{2\sqrt{3}}{\left(\frac{\tilde\alpha(\tilde x)}{2^{2/3}}-\frac{2^{2/3}\tilde x^2}{\tilde\alpha(\tilde x)} \right)}I^2(\tilde x) \nonumber\\
   &\approx& 0.054886.
\end{eqnarray}

\subsubsection{Usadel Equation}
In dirty superconductors, when $\Delta\tau_{\mathrm{el}} \ll 1$, the density of states may be evaluated using the Usadel equation~\cite{usadel_generalized_1970}, which has the following form for  the retarded Green's function
\begin{eqnarray}
\label{eq:RUsadel}
    [\hat{\tau}_3 \epsilon + \hat{\Delta},\hat{g}^R_s] = i D \bm{\nabla}[\hat{g}^R_s \bm{\nabla} \hat{g}^R_s].
\end{eqnarray}
Here $[\  ,\ ]$ denotes the commutator and the hat indicates $2\times 2$ matrices in Gor'kov-Nambu space, which are given by
\begin{eqnarray}
\label{eq:Definitions}
    \hat{\tau}_3 &=& \left(
   \begin{array}{cc}
       1 & 0\\
        0 & -1
        \end{array}
        \right),\ \
        \hat{\Delta} = \left(
   \begin{array}{cc}
         0 &  \Delta \\
        \Delta^* & 0
        \end{array}
        \right),\ \ \\
        \hat{g}_s &=& \left(
   \begin{array}{cc}
       g^R_s & F^R_s\\
        F^{R*}_s & g^{R*}_s
        \end{array}
        \right).
\end{eqnarray}
The Green's function satisfies the nonlinear constraint $\hat{g}^R_s \cdot \hat{g}^R_s  = 1$, and can be expressed in terms of the angles $\theta$ and $\chi$ in the form  $g^R_s = \cos\theta$, $F^R_s = e^{i\chi}\sin\theta$. The density of states is given by
\begin{equation}
    \label{eq:UsadelDOS1}
    \nu(\epsilon)=  \nu_n\Re{\cos\theta(E)}
\end{equation}
and Eq.~\eqref{eq:RUsadel} reduces to
\small
\begin{subequations}
\begin{eqnarray}
   \label{eq:Usadel1}
    \frac{D}{2} \left(   p^2_\mathrm{s} \sin\theta \cos\theta -\nabla^2 \theta \right)
     &=&    i \epsilon \sin\theta +  \Delta \cos\theta , \\
    \label{eq:Usadel2}
     D\bm{\nabla} \left(\bm{p}_s \sin^2 \theta \right)  &=& 0.
\end{eqnarray}
\end{subequations}
\normalsize
Here $\bm{p}_s = \bm{\nabla}\chi$.  As we are considering a thin film we assume $\bm{p}_s$ to be spatially uniform. Then Eq.~\eqref{eq:Usadel2} yields $\bm{\nabla} \theta = 0$, while Eq.~\eqref{eq:Usadel1} reduces to
\begin{eqnarray}
\label{eq:Depairing_Energy}
    \epsilon+i \Gamma \cos\theta = i \Delta \cot\theta
\end{eqnarray}
with $\Gamma = Dp_s^2/2$.

Defining $\xi \equiv e^{i\theta}$ we can write Eq.~\eqref{eq:Depairing_Energy} as a polynomial in $\xi$
\begin{equation}
\label{eq:UsadelCubic}
     2-\xi^{-2}w - i \frac{\eta}{2}\xi^{-3} = 0
\end{equation}
where $w \equiv (\epsilon-\Delta)/\Delta$ as before, $\eta \equiv \Gamma/\Delta$, and we have used the fact that we are concerned only with the energy range $|w| \ll 1$, which corresponds to $|\xi| \ll 1$.
The density of states in this approximation becomes
\begin{equation}
    \label{eq:UsadelDOS2}
    \nu(\epsilon) = \frac{\nu_n}{2} \Re \xi^{-1}.
\end{equation}
One can easily see that the substitution $\xi = i x/\sqrt{2}$ and $\eta = \frac{2\zeta^2}{3\gamma}$  renders Eqs.~\eqref{eq:UsadelCubic} and \eqref{eq:UsadelDOS2}  identical to Eqs.~\eqref{eq:CubicEqnA} and \eqref{eq:IntermediateDOSA}.


\section{
Derivation of the nonlinear conductivity in the absence of \emph{dc} supercurrent}

The microwave power absorbed per unit volume of the sample is expressed in Eq.~(17) in terms of the time-dependence of the energy level $\epsilon(N,t)$. The latter is determined by (c.f. Eq.~(15))
\begin{equation}\label{eq:epsilon_N}
  N = \int_{0}^{\epsilon (N, t)} d \varepsilon \, \nu (\varepsilon, t).
\end{equation}

Although the time dependence of the condensate momentum in the presence of the microwave field is very simple, $p_\mathrm{s}(t) = \frac{e E_\omega}{\omega} \sin(\omega t)$, the density of states $\nu (\varepsilon, t)$, being a nonlinear function of  $p_\mathrm{s}(t)$, is a complicated function of time in the presence of a microwave field. As a result $\epsilon (N, t)$ given by the solution of Eq.~\eqref{eq:epsilon_N} has a complicated dependence not only on time but also on the amplitude of the microwave field, $E_\omega$. Because of  this the absorbed power in Eq.~(17) and the nonlinear conductivity are complicated functions of $E_\omega$ and $\omega$.
The situation simplifies dramatically in the limiting regimes $v_F e E_\omega \Delta \tau_{el}^2/\omega \gg 1$,  and  $v_F e E_\omega \Delta \tau_{el}^2/\omega \ll 1$ . In these regimes the nonlinear conductivity has a simple power law dependence on $E_\omega$ and $\omega$.

\subsection{Ballisic Regime}

For $v_F e E_\omega \Delta \tau_{el}^2/\omega \gg 1$  the quasiparticles contributing to the Debye conductivity are in the ballistic regime and the density of states may be described by Eq.~\eqref{eq:BallisticDOS}. In this case the time-dependent width of the energy window in which the density of states is affected by microwave radiation is  $\delta \epsilon (t) \sim v_F p_\mathrm{s}(t)$.  The characteristic density of quasiparticle levels within the energy window where $\nu (\varepsilon, t)$ is affected by microwave radiation may be estimated as $\delta N \sim \nu_n \sqrt{v_F e E_\omega\Delta / \omega}$. Therefore it is convenient to introduce a rescaled level density
\begin{equation}\label{eq:N_tilde_b}
   \mathcal{N}_b    = \frac{ N}{\nu_n\Delta} \sqrt{\frac{ \omega \Delta }{v_F e E_\omega}}
\end{equation}
and express the time-dependent energy $\epsilon(N,t)$  in terms of a new function $z_{\mathcal{N}_b} (\omega t)$ as
\begin{equation}
\label{eq:CleanEnergyParam}
  \epsilon (N,t)= \Delta +  v_F |p_\mathrm{s}(t) | z_{\mathcal{N}_b}( \omega t).
\end{equation}
Substituting the change of variables \eqref{eq:N_tilde_b}, \eqref{eq:CleanEnergyParam} into Eq.~\eqref{eq:epsilon_N} and using the  form of the density of states in the ballistic regime, Eqs.~\eqref{eq:BallisticDOS} and \eqref{eq:z_eta_b_def}, we find that the dependence of $z_\mathcal{N}( \phi)$ on the phase of the microwave field $\phi=
\omega t$ is determined by the equation
\begin{eqnarray}
  \frac{\mathcal{N}_b}{\sqrt{|\sin(\phi)|}} &=& \frac{\sqrt{2}}{3} \left[\theta (  z_{\mathcal{N}_b}(\phi)+1 )   \left( z_{\mathcal{N}_b}(\phi)+1\right)^{3/2} \right.  \nonumber \\
  &-&  \left. \theta (  z_{\mathcal{N}_b}(\phi)-1 )   \left( z_{\mathcal{N}_b}( \phi)-1\right)^{3/2}\right],
\end{eqnarray}
which does not contain the amplitude $E_\omega$.
Substituting the change of variables \eqref{eq:CleanEnergyParam} into Eq.~(17) we find that the nonlinear conductivity has a simple power law dependence on $E_\omega$ given by Eq.~(18) with the function $F_b(x)$ given by
\begin{widetext}
\begin{equation}\label{eq:F_b}
  F_b(x) = 3\int_0^\infty d\mathcal{N}_b \int_0^\pi \frac{d\phi}{2
  \pi} \int_0^\infty d \bar\tau\   e^{- \bar\tau}
     f_b(\mathcal{N}_b, \phi)f_b(\mathcal{N}_b, \phi - x\bar\tau).
\end{equation}
\end{widetext}
Here $ f_b(\mathcal{N}_b, \phi)  $ denotes the function
\begin{equation}\label{eq:f_b}
  f_b(\mathcal{N}_b, \phi) = \cos(\phi) z_{\mathcal{N}_b} (\phi) + \sin(\phi) \partial_\phi z_{\mathcal{N}_b}(\phi).
\end{equation}
%
%
In the low frequency limit, $\omega\tau_{in} \ll 1$, $F_b(\omega \tau_{\mathrm{in}})$ in Eq.~\eqref{eq:F_b} can be easily evaluated using Eqs.~\eqref{eq:BallisticDOS} and~\eqref{eq:CleanLevelVelocity},
\begin{eqnarray}
    F_b(0) & = & 2\langle \sin^{1/2}(\omega t)\cos^2(\omega t)\rangle  I_b
    \approx 0.108.
\end{eqnarray}
Here $\langle \ldots \rangle $ denotes averaging over half the oscillation period and $I_b$ was defined in Eq.~\eqref{eq:Ib}.

In the high frequency regime $\omega\tau_{in} \gg 1$ we find that $F_b(\omega\tau_{in}) \sim \frac{1}{\omega^2 \tau_{in}^2}$, which can most readily be seen by Fourier transforming Eq.~(17).

\subsection{Diffusive Regime}

At $v_F e E_\omega \Delta \tau_{el}^2/\omega \ll 1$ the quasiparticles contributing to the Debye conductivity are in the diffusive regime,  and the density of states may be described by  Eq.~\eqref{eq:UsadelDOS3}.
In this case the time dependent broadening of the singularity in the quasiparticle density of states is $\delta\epsilon(t) \sim \left( \Delta D^2 p_s^4(t)\right)^{1/3}$. The characteristic density of quasiparticle levels that participate in microwave absorption can then be estimated as $\delta N \sim \nu_n \left(\frac{\Delta^2 D e^2 E_\omega^2}{ \omega^2}\right)^{1/3}$. Introducing a rescaled level density
\begin{eqnarray}
\label{eq:DirtyNTilde}
    \mathcal{N}_d = \frac{1}{\nu_n\Delta} \left(\frac{2\Delta \omega^2}{e^2 D E_\omega^2} \right)^{1/3} N
\end{eqnarray}
we express the time-dependent energy $\epsilon(N,t)$ in terms of a new function $\tilde w_{\mathcal{N}_b}(\omega t)$ as
\begin{equation}
\label{eq:DirtyEnergyParam}
     \epsilon(N,t) = \Delta + \left( \frac{\Delta D^2 p_s^4(t)}{2}\right)^{1/3}\tilde w_{\mathcal{N}_d}(\omega t),
\end{equation}
Substituting Eqs.~\eqref{eq:DirtyNTilde} and~\eqref{eq:DirtyEnergyParam} into~\eqref{eq:epsilon_N} and using the form of the density of states in the diffusive regime, Eq.~\eqref{eq:UsadelDOS3}, we find the dependence of $\tilde w_{\mathcal{N}_d}(\phi)$ on the phase $\phi = \omega t $ is determined by
\begin{equation}
\label{eq:DiffusiveLevelDensity}
    \frac{\mathcal{ N}_d}{\sin^{2/3}(\phi)} =  \int_0^{\tilde w_{\mathcal{N}_d}(\phi)} d\tilde w' \tilde\nu_d(\tilde w'),
\end{equation}
where $\tilde \nu_d(\tilde w')$ is defined by Eq.~\eqref{eq:nutildedDef}. Note that~\eqref{eq:DiffusiveLevelDensity} is independent of the amplitude of the applied field $E_\omega$. Substituting Eqs.~\eqref{eq:DirtyNTilde} and~\eqref{eq:DirtyEnergyParam} into Eq. (17) we obtain Eq. (18) with the function $F_d(x)$ given by
%
\begin{widetext}
\begin{eqnarray}\label{eq:F_d}
    F_d(x) &=& \frac{1}{2^{2/3}\pi}\int_0^\infty d\mathcal{N}_d\int_0^\pi d\phi \int_0^\infty d\bar\tau\ e^{-\bar\tau} f_d(\mathcal{N}_d,\phi) f_d(\mathcal{N}_d,\phi - x \bar \tau)
\end{eqnarray}
\end{widetext}
Here the function $f_d(\mathcal{N}_d,\phi)$ is defined by
\begin{eqnarray}\label{eq:f_d}
  f_d(\mathcal{N}_d,\phi) &=& \frac{2}{3}\sin^{1/3}(\phi) \cos(\phi) \tilde w_{\mathcal{N}_d}(\phi) \nonumber \\ 
  && +\frac{1}{2}\sin^{4/3}(\phi) \partial_\phi\tilde w_{\mathcal{N}_d}(\phi).
\end{eqnarray}

In the low frequency regime, $\omega\tau_{in} \ll 1$, $F_d(\omega\tau_{\mathrm{in}})$ can be easily evaluated. Using Eqs.~\eqref{eq:UsadelDOS3} and~\eqref{eq:UsadelLevelVelocity} we obtain 
\begin{equation}
    F_d(0) = 2 <\sin^{4/3}(\omega t)\cos^2(\omega t)> I_d
    \approx 0.109,
\end{equation}
where  $\langle \ldots \rangle $ denotes averaging over half the oscillation period, and $I_d$ is given by Eq.~\eqref{eq:Id}. In the high frequency limit, $\omega\tau_{in} \gg 1$, $F_d(\omega \tau_{in}) \sim \frac{1}{\omega^2\tau_{in}^2}$ which is most easily seen by Fourier transforming Eq.~(17).